\documentclass[reqno]{amsart}
\usepackage{amssymb, latexsym}

\begin{document}

\title{On Analytic Model for Airflow around $2$-Dimensional Composite Airfoil}
\author{ William B. Ribbens$^1$ and   Rita Gitik$^2$ \\ University of Michigan, Ann Arbor, Michican, 48109, USA}

\begin{abstract}
This paper presents a method of computing section lift characteristics for a $2$-dimensional airfoil with a second $2$-dimensional object at a position
at or ahead of the leading edge of the airfoil. Since both objects are $2$-dimensional, the analysis yields a closed form solution to calculation of the airflow over the airfoil and second object, using conformal mapping of analytically closed form airflow velocity vector past two circular shaped objects in initial complex plane, using a standard airflow model for each object individually. The combined airflow velocity vector is obtained by linear superposition of the velocity vector for the two objects, computed individually. The lift characteristics are obtained from the circulation around the airfoil and second object which is computed from the combined closed form velocity vector and the geometry along the contour integral for circulation.
The illustrative example considered in this paper shows that the second object which is essentially a cylinder whose diameter is approximately $9\%$ of 
the chord length of the airfoil reduces the section lift coefficient by approximately $6.3\%$ from that of the airfoil alone.
\end{abstract}

\date{\today}

\footnote{Professor Emeritus, Department of Engineering and Computer Science and Professor Emeritus, Department of Aerospace Engineering.}
\footnote{Visiting Scholar, Department of Mathematics.}

\maketitle

\section{Introduction}

Precise calculations of airflow past an airplane or other aircraft structures
are accomplished with CFD. Computational results are in the form of numerical
samples at discrete locations around the surfaces of the aircraft.
However, CFD computations do not yield a closed form analytic model for such airflow.

Closed form analytic solutions to airflow is possible for $2D$ models.

One method is a conformal mapping. Traditional applications map a circle or other simple shape
in a complex plane $z=x+iy$ to an airfoil shape in the conformal map 
complex plane $\eta = \epsilon + i \nu$, e.g. $\eta = z + \frac{a^2} {z}$, see for example \cite{Ha} and \cite{Iv}.

This paper presents a closed form solution to a $2D$ model of airflow around a pair
of objects.The objects consist of a primary circle in the $z$-plane which maps to an
airfoil in the $\eta$-plane. The second object is a circle which is much smaller than
the primary and is upstream of the primary circle in an incident uniform airflow field.
This secondary object maps to a relatively small ellipse with near unity ratio of major to minor axis, in contact with the
airflow at its leading edge with the diameter along the airfoil cord.

With respect to the airflow over winglike structures, the conformal map of the two $z$-plane circles can be interpreted as the
intersection of a wing with a leading edge object, both having infinite dimensions orthogonal to the $\eta$-plane.
This is the interpretation of traditional $2D$ airfoil representations. 
For the present analysis the airflow around the $2$ mapped structures in the 
$\eta$-plane are taken to be the superposition of the conformal mapping of the individual
airflow around the $2$ objects in the $z$-plane.

The present analysis assumes that the incident airflow on the primary surface in the
$z$-plane is a uniform flow field at an angle of attack of $\alpha_0$. 
This assumption is equivalent to ignoring the influence of the secondary object
on the flow field of the primary.

A model for the velocity vector of the airflow around each of the two circular objects in the $z$-plane
at any location relative to the geometry of the object including two coordinates of the center of each circle has been known for a long time.
For the purpose of the present paper, the airflow around a concentric circle of the first object due to the influence of the second
object is computed. The velocity vector of interest is the superposition of the airflow due to the second object with the model for the air velocity
along the same contour of the first object. Conformal mapping of the combined velocity along the contour to the $\eta$-plane yields a model for the air velocity vector from which circulation $\Gamma$ can be computed. This circulation provides the basis for computing the section lift coefficient of the two objects in the $\eta$-plane.

\section{Airflow around One Object}

The closed form calculation of inviscid airflow around an object in terms of tabulated functions (e.g. $sin$, $cos$) requires that the object have separable boundary conditions in the corresponding coordinate system. The airflow is derived from a potential function which is here denoted $\phi$ and an associated stream flow function, here denoted $\psi$. Both functions satisfy Laplaces equation, as explained in \cite{Ku} p.81:

$\bigtriangledown^2 \phi =0$

$\bigtriangledown^2 \psi =0$.

An example of an object satisfying the above conditions for a closed form solution in
$3$ dimensions is a sphere. However, a sphere is not a useful shape for application to aircraft airflow analyses.
 
On the other hand, an object having potential application for closed form airflow is a cylinder with an infinitely long axis. This hypothetical shape can be meaningfully interpreted as a $2$-dimensional
representation as has, of course, been used for a very long time to asses sectional wing airfoil characteristics.

One of the well-known analytic  methods for calculating the equivalent of airflow over a $2$-dimensional airfoil shape has involved conformal mapping of a circle in one complex plane to an airfoil shape in another complex plane. This airfoil shape can be made to have a contour with a rounded leading edge and a sharp trailing edge. By proper selection of the location of the circle in the original complex plane relative to the coordinate origin, 
the conformal mapping yields an airfoil shape with a desired thickness and camber.

For the purposes of the present paper, the second object is a circle in the $z$-plane, having a radius that is much less than that of the first object.
The origin of the second object is at $z=x_{cc}+iy_{cc}$, where $x_{cc} > 0$, $y_{cc}=0$. The second object is mapped to the $\eta$-complex plane
via a similar transformation to the first object. The mapped second object is a small ellipse which is outside the airfoil but is tangent to the leading edge.

The incident airflow is uniform with a velocity at $\infty$ of $V_0$ and at an angle $\alpha_0$ to the $x$-axis.
This incident airflow into $\eta$-plane with the equivalent of a positive angle of attack (AOA) of $\alpha_0$ on the mapped airfoil.
The object of the present paper is to calculate the airflow around the surface of the mapped airfoil in the $\eta$-plane due to the influence of both
objects. Since the method for the airflow is linear, the combined airflow is obtained by superposition of the airflow along the airfoil due to each object
separately.

The $z$-plane complex variable is denoted

$$z=x + iy$$

where $x$ is the real axis and $iy$ is the imaginary axis. The other complex variable is denoted $\eta$,

$$\eta = \epsilon + i\nu$$

where $\epsilon$ is the real axis and $i\nu$ is the imaginary axis.

The conformal mapping from $z$ to $\eta$ is of the generic form

\begin{equation}
 \eta = z + \frac{a^2}{z}
\end{equation}
where $a$ is a real parameter. 

The airflow local velocity in either complex plane is derived from a complex function which is denoted $W$ and is given by

$$W=\phi + i \psi$$

where $\phi$ is velocity potential function and $\psi$ is a stream function.

Both of these functions satisfy Laplaces equation for irrotational flow:

$$\bigtriangledown^2 \phi =0$$
$$\bigtriangledown^2 \psi =0$$

For simplicity of analysis, it is assumed that the object $1$ circle center is on the $x$-axis at $x_c$.
The conformal map of this circle yields a symmetric airflow.
It has long been known that the complex potential function for uniform flow at and angle $\alpha_0$ with velocity
at infinity at $V_0$ and with circulation $\Gamma$ about a cylinder with center at $x_c$ is given by

\begin{equation}
W= V_0 \left[ (z-x_c)e^{-i\alpha_0} + \frac{(a- x_c)^2 e^{i \alpha_0}}{(z-x_c)} \right]+
 \frac{i \Gamma}{2 \pi} ln \left[ \frac{(z-x_c)e^{-i\alpha_0}}{(a-x_c)} \right]
\end{equation}

It is also well known, see for example \cite{A-V} p.50, that the air velocity in the $z$-plane $(V(z))$ is given by:
$$V(z) = \frac{dW}{dz}$$
\begin{equation}
= V_0\left[(e^{-i\alpha_0} - \frac{(a - x_c)^2}{(z -x_c)^2} e^{i \alpha_0}\right] + 
\frac{i \Gamma}{2 \pi(z-x_c)}
\end{equation}

Similarly the air velocity in the $\eta$-plane is given by
$$V(\eta) = \frac{dW}{d \eta}$$
$$=  \frac{dW}{dz} \frac{dz}{d \eta}$$
\begin{equation}
=V(z) \frac{dz}{d \eta}
\end{equation}
The latter factor $\frac{dz}{d \eta}$ is given by
$$\frac{dz}{d \eta}= \frac{1}{\frac{d \eta}{dz}}$$
\begin{equation}
=\frac{z^2}{z^2-a^2}
\end{equation}

Substituting $V(z)$ from equation $3$ and $\frac{dz}{d \eta}$ from equation $5$, yields the following expression for $V(\eta)$.
\begin{equation}
V(\eta)=\left[  V_0 \left(e^{-i\alpha_0} - \frac{(a - x_c)^2 e^{i \alpha_0}}{(z-x_c)^2} \right)+
 \frac{i \Gamma}{2 \pi(z - x_c)} \right]\left(\frac{z^2}{z^2-a^2} \right)
\end{equation}

It is also well known that $z=a$ corresponds to the trailing edge of the airfoil in the $\eta$-plane. The velocity
at the trailing edge would be infinite unless the first factor $V(z)$ at $z=a$ is zero, as shown in \cite{A-V} p.52. That is $V(z)|_{z=a}=0$.
This condition is satisfied for a specific value for the circulation $\Gamma$ which results from the well known Kutta-Joukowski condition.
This value for $\Gamma$ is given by:
\begin{equation}
\Gamma = 4 \pi (a -x_c)V_0sin(\alpha_0)
\end{equation}

One of the important variables in the application of airflow computation via conformal mapping is the section lift which is denoted $l$.
This is technically equivalent to the lift per unit span length of the theoretical infinitely span wing which has an airfoil given by the  
conformal mapping of the first object. The section lift can be computed directly from the circulation $\Gamma$ around the airfoil from the following relationship:
\begin{equation}
l=\rho V_0 \Gamma
\end{equation}

where $\rho =$air density.

In addition the circulation can be computed from the velocity vector along streamlines with the following formula:

\begin{equation}
\Gamma=\oint_{c_s} \overline{V}\cdot \overline{ds}
\end{equation}

where $c_s=$closed contour around the airfoil along streamlines,

$\overline{V}=$vector notation for the velocity,
 
$\overline{ds} =$differential length vector along $c_s$.

As is also well known, the section lift was characterized by a section lift coefficient $c_l$ as given below:
\begin{equation}
l=qSC_l
\end{equation}
 
where 

$q=\frac{1}{2} \rho V_0^2=$dynamic pressure,

$S=$section area per unit span length$=c=$section chord length.

The section lift coefficient $C_l$ is computed using $\Gamma$ as given by
$$C_l=\frac{l}{Sq}$$
$$=\frac{2\rho V_0 \Gamma}{\rho V_0^2c}$$
\begin{equation}
=\frac{2\Gamma}{V_0c}
\end{equation}

For the purpose of evaluating the present method of calculating a closed form solution to the $2$ object airflow, a numerical calculation of
$\Gamma$ and $C_l$ was developed using the closed form solution. As a means of assessing the numerical calculation of $\Gamma$, the procedure was first applied to the first single object in which the $z$-plane circle maps to the airfoil.

The contour integral for $\Gamma$ in terms of the complex variables can be formulated as the contour integral of a scalar complex variable. This interpretation utilizes the following notation:

$$V(\eta)=u+iv$$
$$ds=d\epsilon +i d\nu$$

The dot product  $\overline{V}\cdot \overline{ds}$ can be expressed as the following:

$$\overline{V(\eta)} \cdot \overline{ds} = u d\epsilon + v d\nu = Re[V(\eta)d\eta^*]$$

Thus, the contour integral of a vector product becomes the following

\begin{equation}
\Gamma=\oint_{c_s} \overline{V}\cdot \overline{ds}= \oint_{c_s}Re[V(\eta)d\eta^*]
\end{equation}

The numerical evaluation of $\Gamma$ is accomplished by computing the airflow velocity $V(\eta)$ corresponding to the conformal mapping of the airflow
velocity at $K$ points around the circle in the $z$-plane.

The algorithm for calculating $V(\eta_k)$ is based upon equation $6$ evaluated at $K$ points $z(k)$ which are computed as follows:

for $k = 1:K$

$\theta(k)=\frac{2 \pi k}{K}$

$x(k)=a_1 cos(\theta(k)) +x_c$

$y(k)=a_1 sin(\theta(k))+y_c$

$z(k) = x(k) + iy(k)$

where

$a_1=1.2$

$x_c=0.200$

$y_c = 0$

The conformal mapping from $z(k)$ to $\eta(k)$ is given by
$$\eta(k)=z(k) + \frac{a^2}{z(k)}$$
where $a=1.4306$.

The differential complex length $\delta s(k)$ in the formula for $\Gamma$ is computed as follows:
\begin{equation}
s(n) = \eta(k) - \eta(k-1), k=1,2, \cdots , K
\end{equation}
where $\eta(0)=2a$.

For sufficiently large $K$ the contour integral for $\Gamma$ is closely approximated by the following summation:

$$\Gamma = \oint_{c_s} V \cdot ds$$
\begin{equation}
\cong \sum_{k=1}^K Re[V(\eta(k)) \delta s^*(k)]
\end{equation}

A computation was made for a specific example with the following parameters:

\begin{center} $V_0=169ft/sec$ (i.e. $100Kt$)
\end{center}
$$\alpha_0=0.0931rad$$
$$K=250$$

Using the Kutta-Joukowski condition in the form of equation $7$, the circulation is given by:
$$\Gamma=252.02$$

The numerical evaluation of equation $14$ yields a value for $\Gamma$:
$$\Gamma=252.25$$

The relatively close agreement between the two calculations for $\Gamma$, which ideally should be identical, indicates that the
$K$ used is sufficient for the purposes of the present paper. The lift coefficient, computed using equation $11$ is given by
$$C_l=0.5362$$

A similar calculation for $\alpha_0 =0$ yields $C_l=0.0031$.

The slope of the section lift coefficient which is denoted $C_{l \alpha}$ is given by
\begin{equation}
C_{l \alpha}=\frac{\partial C_l}{\partial \alpha}=\frac{0.5362-0.0031}{0.0931}=5.72
\end{equation}

The value for $c_{l \alpha}$ is consistent  with many geometrically similar airfoils, as published by NACA during the post WWII years.

\section{Airflow around Two Objects}

For an understanding of the present method it is, perhaps, helpful to review the geometry of the two objects in both the $z$-plane
and the $\eta$-plane. 

The second object is a circle in the $z$-plane which yields an ellipse at the leading edge of the airfoil with the conformal map defined above.
The ratio of the major to minor axes of this ellipse is $1.045$ which is very close to that ratio for a circle, which is $1$.

The method of calculating the airflow around the airfoil which is created by conformal mapping the first object circle due to the influence of the second object begins with calculation in the $z$-plane. For this $z$-plane calculation, a new coordinate system, which is denoted $z_2$, with an origin at the center of the second object center is given by:
\begin{equation}
z_2=z-x_{cc}
\end{equation}

With this coordinate system the airflow at any point in the $z$-plane can be derived from the potential function, which is denoted $W_2$.

For this function a uniform airflow with velocity $V_0$ at infinity, which is at angle $\alpha_0$ to the $x$-axis at the second object, the function
$W_2(z_2)$ is given by 

\begin{equation}
W_2(z_2)=V_0 \left( z_2e^{-i\alpha_0} + \frac{a^z_c e^{-i\alpha_0}}{z_2} \right) + i\Gamma_2 ln\left(\frac{z_2}{a_c}e^{-i\alpha_0}\right)
\end{equation}
where $a_c=$radius of the second object circle

$\Gamma_2=$circulation about the second object.

The velocity of any point $z_2$ in the $z$-plane, which is denoted $V_2(z_2)$, is given by:
$$V_2(z_2)=\frac{dW_2(z_2)}{dz_2}$$
\begin{equation}
=V_0\left( e^{-i\alpha_0}-\frac{a_c^2}{z_2^2}e^{-i\alpha_0}\right) + \frac{i\Gamma_2}{2\pi z_2}
\end{equation}

The corresponding velocity in the $z$-plane as a function of the original complex coordinate $z$ is denoted $V_2(z)$ and is given by:
\begin{equation}
V_2(z)=\frac{dW_2(z_2)}{dz_2} \frac{dz}{dz_2}
\end{equation}
where the second factor on the right hand side is unity. The resulting $V_2(z)$ is given by:
\begin{equation}
V_2(z)=V_0\left(e^{-i\alpha_0}-\frac{a_c^2e^{-i\alpha_c}}{(z-x_{cc})^2}\right) + \frac{i\Gamma_2}{2\pi(z-x_{cc})}
\end{equation}

The calculation of the velocity component along the airfoil or along contour $c_s$ in the $\eta$-plane is based upon calculation of $V_2(z)$ for the same $z_k$ points used in the calculation of $V(\eta)$ for the first object. This calculation yields the velocity vector $V_2(\eta_k)$ which is given by
$$V_2(\eta_k)=V_2(z_k)\frac{dz}{d\eta}$$
\begin{equation}
=V_2(z_k)\left(\frac{z_k^2}{(z_k^2-a^2)}\right)
\end{equation}

As in the case of the velocity $V(\eta_k)$ for the first object, the requirement of a finite value for $V_2(\eta_k)|_{z_k=a}$ is stated in the condition:
$$V_2(z_k)|_{z_k=a}=0$$

This latter condition requires a specific value for $\Gamma_2|_{z_k=a}$ which is given by:

\begin{equation}
\Gamma_2(a)=2\pi i(a-x_{cc})V_0\left[e^{-i\alpha_0}-\frac{a_c^2}{(a-x_{cc})^2}e^{i\alpha_0}\right]
\end{equation}

The above value for $\Gamma_2$ is used in equation $20$ to calculate $V_2(z)$ and the result of this calculation about the airfoil which is denoted 
$\Gamma_2$ is computed using the following modified version of equation $14$:

$$\Gamma_2= \sum_{k=1}^K Re[V_2(\eta(k)) \delta s^*(k)]$$
where $\delta s(k)$ is defined in equation $14$.

The total circulation, which is denoted $\Gamma_T$ is the linear superposition of $\Gamma$ and $\Gamma_2$:
$$\Gamma_T=\Gamma+\Gamma_2$$
where $\Gamma$ is defined in equation $14$.

The section lift coefficient for the combined $2$ objects, which is denoted $C_{lT}$, can be obtained using equation $11$ with $\Gamma_T$ substituted for $\Gamma$.
$$C_{lT}=\frac{2\Gamma_T}{V_0c}$$

For the specific parameters of the objects, including size and location, and for the incident air velocity $V_0$ at angle $\alpha_0$ to the $x$-axis,
$C_{lT}$ is computed to be:
$$C_{lT}=0.5024$$
This constitutes a reduction in $C_l$ from the single object case $(C_l=0.5362)$ of approximately $6.3\%$.
 
The combined lift coefficient for $\alpha_0=0$ is denoted $C_{lT}(0)$ which is computed to be
$$C_{lT}(0)=-0.0101$$

The lift slope for the combined objects, which is denoted $C_{lT\alpha}$, is given by:

$$C_{lT\alpha}=\frac{\partial C_{lT}}{\partial\alpha}=\frac{0.5024+0.0101}{0.0931}=5.475$$

Thus, the influence of the second object on the section lift coefficient is to reduce $C_{lT}$ as well as to reduce the section lift coefficient slope by about $4.5\%$.

\section{Conclusions}

In principle, the method of this paper could be used to calculate in closed form the air velocity vector around an airfoil with a leading edge object.
However, the second object in the $z$-plane must be a circle to have a closed form solution for the airflow. The shape of the leading edge object is determined by the size and location of the second object circle. Thus, in practice, only certain leading edge shapes can be generated by conformal mapping of the second object circle.

\end{document}